# Enhancing Synchrony in Chaotic Oscillators by Dynamic Relaying

Ranjib Banerjee[1,3], Dibakar Ghosh[2], E.Padmanaban[3], R.Ramaswamy[4], L.M.Pecora[5], Syamal K.Dana[3]
[1]Department of Mathematics, Gargi Memorial Institute of Technology, Kolkata, India
[2]Department of Mathematics, Dinabandhu Andrews College, Kolkata 700084, India
[3]Central Instrumentation, CSIR-Indian Institute of Chemical Biology, Kolkata 700032, India
[4]University of Hyderabad, Hyderabad 500 046, Hyderabad, India
[5]Code 6340, Naval Research Laboratory, Washington, D.C. 20375, USA

In a chain of mutually coupled oscillators, the coupling threshold for synchronization between the outermost identical oscillators decreases when a type of impurity (in terms of parameter mismatch) is introduced in the inner oscillator(s). The outer oscillators interact indirectly via dynamic relaying, mediated by the inner oscillator(s). We confirm this enhancing of critical coupling in the chaotic regimes of Rössler system in absence of coupling delay and in Mackey-Glass system with delay coupling. The enhancing effect is experimentally verified in electronic circuit of Rössler oscillators.

Chaotic trajectories in coupled nonlinear dynamical systems are known to synchronize when the strength of the coupling exceeds a critical value [1-3]. Such complete synchronization (CS) occurs only when the oscillators are identical. When the oscillators are mismatched, CS becomes unstable due to attractor bubbling or bursting instabilities [4], but lag synchronization (LS) may arise at a lower value of the critical coupling [5, 6]. An unexpected consequence of this lag synchrony is that it can promote complete synchrony in a chain of diffusively coupled chaotic oscillators when there are isolated *impurities*, namely, mismatched oscillators. The simplest case arises when there are three oscillators coupled as shown in Fig. 1(a), with oscillators 1 and 3 being identical to each other but mismatched with the oscillator 2. The central oscillator with either positive or negative mismatch induces a common time lag or lead with the outer oscillators, leading to a LS scenario at a lower critical coupling. Examination of the master stability function (MSF) [7] in Fig. 1(b) for the outer system in a set of three chaotic Lorenz systems shows that the CS occurs at a lower critical coupling for mismatch in some regions of the parameter space of the central oscillator. This enhancing effect is seen at the right side of the dark hill where the coupled dynamics is chaotic. On the left side, the coupled dynamics is periodic where a diminishing effect is observed instead. The MSF is calculated between (1, 3)-pair of identical oscillators which are not directly coupled but interact through exchange of signals mediated by the oscillator 2. The effect is found to occur both for instantaneous and delay coupled systems.

In presence of long conduction delays, zero-lag synchronization (ZLS) was reported [8-9] in two distantly located populations of neurons in cerebral cortical areas when mediated by a third population. The ZLS was verified in laboratory experiments on laser [10] and electronic circuit [11]. In these experiments, one common strategy was to detune (positive or negative) the frequency of the central oscillator to optimize the ZLS. The robustness of the ZLS to frequency detuning in the central oscillator [12-13] and also to parameter mismatch in the outer oscillators [13] was investigated. An isochronal synchronization [14] was also reported in three instantaneously coupled laser sources. The central oscillator there played a leader/laggard role [15-16] in presence of long coupling delay. However, the role of parameter mismatch (or frequency detuning) on the critical coupling of ZLS has not been investigated. In this Brief Report, we address a natural question: does the parameter mismatch in the central oscillator play any role on the ZLS or isochronal synchronization? Investigating a chain of three delay coupled delay Mackey-Glass systems with coupling delay, we find that two identical outer oscillators evolves to a CS state at a lower critical coupling for parameter mismatch in the central oscillator: this is defined as enhancing of synchrony in the outer oscillators.

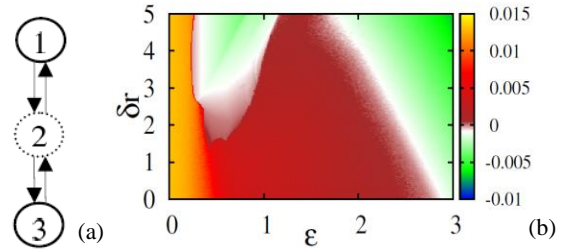

FIG.1. (color online) MSF ($\lambda_{max}$) as a function of coupling and mismatch. Each circle in (a) denotes a Lorenz system: $\dot{x}_i = \sigma(y_i - x_i)$, $\dot{y}_i = r_i x_i - y_i - x_i z_i$, $\dot{z}_i = -bz_i + x_i y_i$ ($i$=1,2,3) : $b$=8/3, $\sigma$=10, $r_1$=$r_3$=28, $r_2$=28+$\delta r$, $\delta r$ is a mismatch and coupling strength $\varepsilon$. Arrows denote mutual coupling via $y$-variable. $\lambda_{max}$ plot (b) has two CS regions (green-white) delineated by critical coupling boundary on both sides of a hill when $\lambda_{max}$ crosses positive (red) to a negative value (green-white). Slope of critical coupling boundary is negative at right while positive at left.

The coupled Mackey-Glass is given by,

$$\dot{x}_i = -a\, x_i + \frac{m_i\, x_i(t-\tau_0)}{1+x_i^{10}(t-\tau_0)} \qquad i=1,2,3 \qquad (1)$$
$$+ \varepsilon(x_{i-1}(t-\tau_1) + x_{i+1}(t-\tau_1) - 2x_i)$$

with zero flux $(x_0 \equiv x_1, x_4 = x_3)$. $a_i$, $m_i$ are constants, $\tau_0$ is the intrinsic delay of the system. A mismatch, $\delta m$ ($m_2 = m_{1,3} \pm \delta m$), is introduced in the central oscillator. We consider a large coupling delay $\tau_1 = 4$ with a coupling strength $\varepsilon$ to enact the long range ZLS [8-11]. Numerically computed cross-correlation $\rho_{x_1 x_3}$ [17] of the ($x_1$, $x_3$)-pair of time series of the outer oscillators is plotted in the $\varepsilon$-$\delta m$ plane in Fig. 2(a). In case of CS, $\rho_{x_1 x_3} = 1$. Figure 2(a) shows a negative slope of the boundary line that separates the CS regime (black) from the non-synchronous regime (grey). The negative slope of the boundary confirms the enhancing effect. The boundary is possibly fractal, and a further exploration of this feature is presently underway. We reconfirm the enhancing effect by plotting $\rho_{x_1 x_3}$ with $\varepsilon$ in Fig. 2(b) for two specific choices of mismatch. The critical coupling decreases with mismatch in the central oscillator. The ($x_1$, $x_3$)-pair of time series in Fig. 3(a) shows poor correlation ($\rho_{x_1, x_3} \approx 0.70$) in Fig. 3(d) at zero lag for $\varepsilon = 0.53$ when all oscillators are identical. Figure 3(b) shows CS between the ($x_1$, $x_3$)-pair of time series for the same $\varepsilon = 0.53$ at zero lag ($\rho_{x_1, x_3} = 1.0$) in Fig. 3(e) when a mismatch ($\delta m = 0.3$) is introduced in the central oscillator. The outer oscillators maintain a time lag with the central one as shown in Fig. 3(c). However, the lag ($\tau = 4.22$) is slightly larger than the coupling delay ($\tau = 4.0$) as shown in the maxima of $\rho_{x_1 x_3}$ in Fig. 3(f). It is checked, by taking coupling delay, $\tau = 0$, that the additional lag appears due to the parameter mismatch. This indicates that a LS scenario leads to enhancing of CS in the outermost oscillators due to a parameter mismatch in the central oscillator.

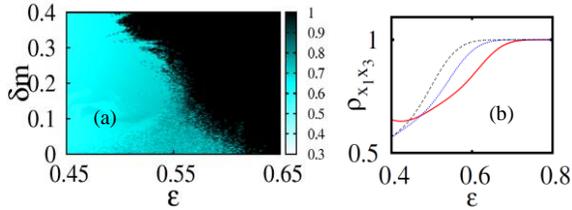

To further elucidate that the LS effect plays a key role in lowering the critical coupling of CS, we consider a chain of three instantaneously coupled Rössler oscillators,

$$\dot{x}_i = -\omega_i y_i - z_i + \varepsilon(x_{i+1} + x_{i-1} - 2x_i)$$
$$\dot{y}_i = \omega_i x_i + a y_i, \quad \dot{z}_i = b + z_i(x_i - c) \quad i = 1, 2, 3 \quad (2)$$

with zero flux $(x_0 \equiv x_1, x_4 = x_3)$. We choose $a = 0.2$, $b = 0.4$, $c = 7.5$, $\omega_1 = \omega_3 = 1$, $\omega_2 = \omega_1 + \delta \omega$, $\delta \omega = 0.1$ for a chaotic regime.

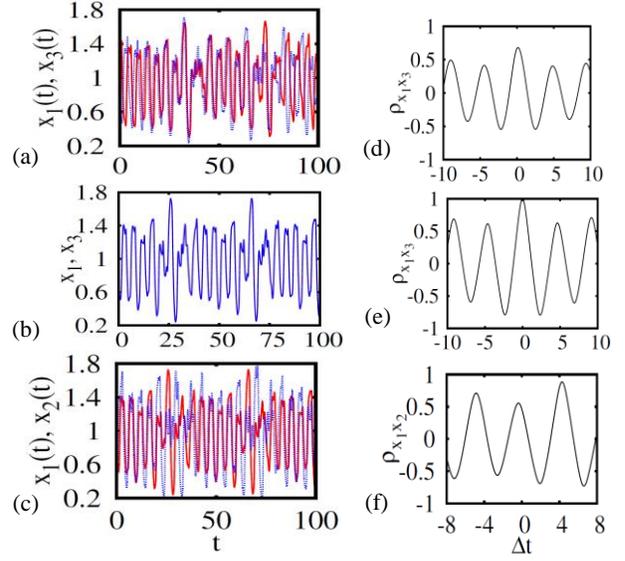

FIG. 3 (color online). Mackey-Glass System: ($x_1$, $x_3$)-pair of time series in black (red) and grey (blue) lines for identical case (a), mismatch case (b), of ($x_1$, $x_2$)-pair in black (red) and grey (blue) lines for mismatch case (c). Corresponding $\rho_{x_1 x_3}$ of ($x_1$, $x_3$)-pair of time series in (d) and (e), of ($x_1$, $x_2$) in (f). $\varepsilon = 0.53$, $a = 1.1$, $m_1 = m_3 = 3$, $\delta m = 0.3$, $\tau_0 = 2.0$, $\tau_1 = 4.0$.

Numerical results in Fig. 4 show that $\lambda_{max}$ crosses from positive to a negative value at a lower critical coupling (solid line) when a mismatch is introduced in the central oscillator. The (1, 3)-pair of oscillators is not in CS in Fig. 5(a) and 5(d) for $\varepsilon = 0.13$ when three oscillators are identical. In fact, they need a larger coupling ($\varepsilon > \varepsilon_c = 0.22$) to develop a CS state. Instead, when a mismatch ($\delta \omega = 0.1$) is introduced in the central oscillator, the (1, 3)-pair of oscillators emerges into a CS state for the lower critical coupling, $\varepsilon_c = 0.13$ as shown in Fig. 5(b) and 5(e). The outer oscillators then emerge into a LS state with the central oscillator as confirmed by the ($x_1$, $x_2$)-pair of time series in Fig. 5(c). A lag ($\tau = 0.5$) is created in the LS state as shown in the $x_1$ vs.$x_2(t+\tau)$ plot in Fig. 5(f). The onset of CS in the outer identical oscillators is thus enhanced by the mismatched central oscillator via dynamic relaying. The lag synchrony lowers the critical coupling [5-6] and eventually enhances CS in the outer oscillators in chaotic regimes of our example systems. On the contrary, a diminishing effect is seen in Fig. 4(b) when the Rössler oscillators are in the limit cycle regime. This also holds in 2D limit cycle systems, which effect is yet to be fully understood.

We experimentally support the enhancing effect using electronic analog of Rössler oscillators (circle) coupled diffusively (square box) as shown in a block diagram (Fig.6). A single Rössler oscillator circuit is only shown with details of the coupling circuit. The resistance $R$ of the outer oscillators (1, 3) is selected as 100kΩ to make them

FIG. 2 (color online). Mackey-Glass system: $\rho_{x_1 x_3}$ plot of ($x_1$, $x_3$)-pair of time series in $\varepsilon - \delta m$ plane (a); CS in black region. Critical coupling boundary delineated by black and grey (blue) regions shows a negative slope. $\rho_{x_1 x_3}$ of ($x_1$, $x_3$)-pair in (b) with $\varepsilon$ for $\delta m = 0$ in black (red), 0.2 in grey (blue), 0.35 in dotted lines. $a = 1$, $m_1 = m_3 = 3$, $\tau_0 = 2$.

closely identical while it is chosen as 91kΩ for the middle oscillator to introduce a mismatch. The resistance $R_C$ is then varied to find a critical coupling (1.8kΩ) when (1, 3)-oscillators are almost in CS as seen in the upper panels of Fig. 7. This leads the outer oscillators to a LS state with the central one (middle panels). Next, keeping $R_C$ unchanged, $R$ in the central oscillator is changed to 100kΩ for making the three oscillators almost identical when they are all desynchronized (lower panels). A larger coupling ($R_C$=2.5kΩ) is found necessary to induce CS in three identical oscillators. Experimental results are found in good agreement with the numerical results in Fig. 5.

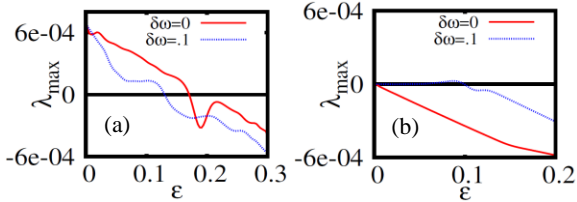

FIG. 4 (color online). MSF ($\lambda_{max}$) of outer Rössler oscillators with $\varepsilon$. In chaotic regime, (a) $\varepsilon=\varepsilon_c=0.22$ for CS in outer oscillators for identical case in black (red) line, $\varepsilon_c=0.13$ for mismatch ($\delta\omega=0.1$) in central oscillator in grey (blue) line. In limit cycle case, (b) $b=0.2$, $c=2.5$ (other parameters same), the $\varepsilon_c$ for CS in outer oscillators is larger for mismatched case in grey (blue) line than the identical case in black (red) line.

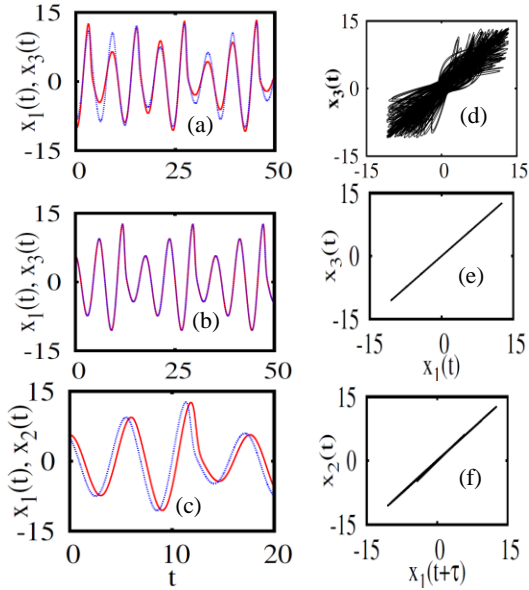

FIG. 5 (color online). Module of three Rössler oscillators. Pair of ($x_1$, $x_3$) times series in black (red) and grey (blue) lines at $\varepsilon=0.13$ in identical case (a), mismatch case (b), of ($x_1$, $x_2$) time series in black (red) and grey (blue) lines in LS state (c). Plot of $x_1$ vs. $x_3(t)$ in identical case (d), mismatch case (e), of $x_1$ vs. $x_2(t+\tau)$ in LS (f) when $\tau=0.5$.

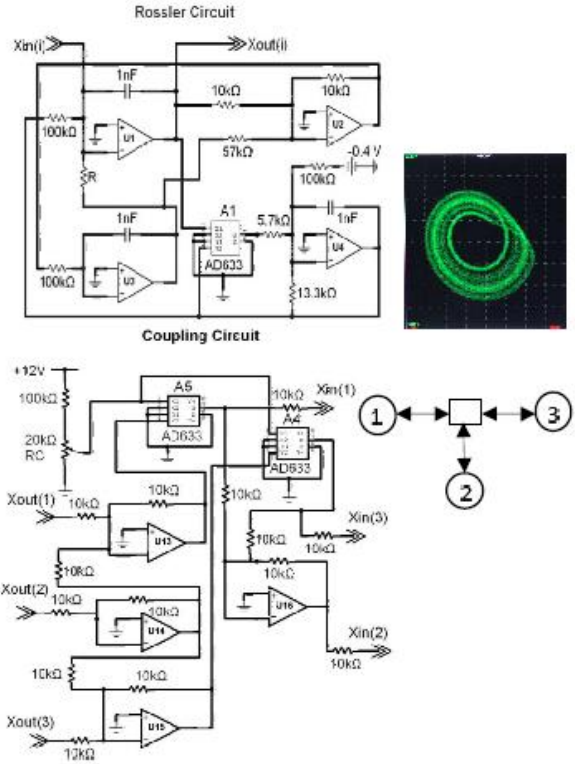

Fig. 6. Rössler oscillator and coupling circuit. Upper circuit is a single Rössler oscillator; attractor of an isolated oscillator at right (oscilloscope picture). Lower circuit derives the diffusive coupling scheme (square box) in three oscillators (circle). $R$ in oscillator (2) is varied to induce mismatch. Coupling strength increases with $R_C$ as a proportional constant voltage.

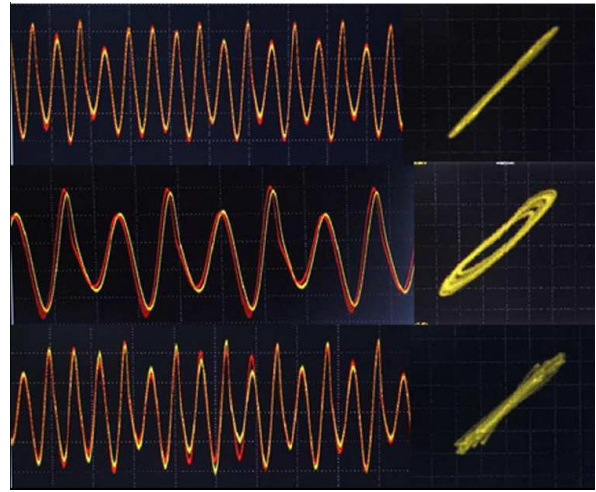

Fig. 7 (color online). Oscilloscope pictures. Left panels: pair of time series of outer oscillators in almost CS in upper row, of the central and one of the outer oscillators in LS in middle row, of outer oscillators in lower row when all are identical. Right panels: phase portraits of similar state variables of two oscillators; each right panel corresponds to its immediate left panel.

The enhancing effect is found in 1D arrays of $N$-oscillator ($N>3$) (Fig. 8) as well. In an array of 4-oscillator, two outermost identical oscillators are mediated by two oscillators identically mismatched. In a second array of 5-oscillator, four mutually coupled identical oscillators are mediated by a mismatched oscillator. The oscillators at symmetric positions on both sides of the central oscillator emerge into CS at a lower critical coupling. In both the cases, $\lambda_{max}$ of the outer two oscillators crosses to a negative value at a lower critical coupling for a mismatch introduced in the intermediate oscillator(s) in Fig. 9.

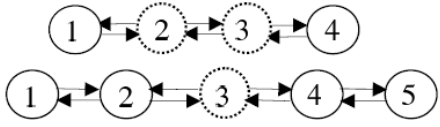

FIG. 8. 1D array of bidirectionally coupled *N*-oscillator: identical (solid circle) and mismatched oscillators (dotted circle). Upper row: identical outer oscillators (1, 4) mediated by 2-identically mismatched oscillators. Lower row: 4-identical oscillators mediated by a mismatched oscillator (3).

Disorder enhanced synchrony was reported earlier [18] in an array of coupled Josephson junctions, as is noise induced enhancing of phase synchronization [19] or coherence resonance [20]. However, the mechanism of enhancing of synchrony reported here is different.

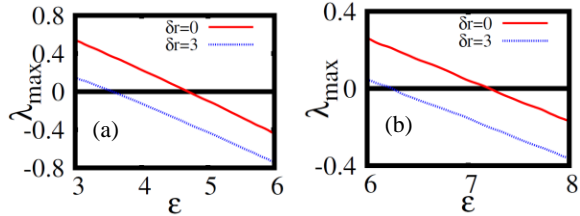

FIG. 9 (color online). MSF ($\lambda_{max}$) of outer oscillators in a chain of *N*-Lorenz oscillators in Fig. 9, (a) *N*=4, (b) *N*=5. Critical coupling of CS in outer oscillators is lower in grey (blue) line for mismatched central oscillator than all identical case in black (red) line for both (a) and (b). $\sigma$=10, *r*=28, *b*=8/3, $\delta r$=3.0.

To summarize, an enhancing of synchrony is reported here in a chain of identical oscillators mediated by mismatched oscillator(s). A common time lag is created between the identical outer oscillators and the mismatched central oscillator leading to a LS scenario at a lower critical coupling. This time lag played a role of dynamic relaying the outer oscillators to establish an indirect coupling between them and thereby enhances CS in the outer oscillators. We presented several example systems to verify the LS scenario causing the enhancing effect both in presence and in absence of coupling delay. We provided experimental evidence using electronic circuit of Rössler oscillators. An enhancing of synchrony was reported earlier [21] in two oscillators by an induced coupling delay when the coupled system switches from a chaotic to a periodic state. In the present instance, the coupled oscillators remain chaotic before and after the coupling. This enhancing effect is found for a negative mismatch too where the central oscillator leads the outer ones instead of lagging. The effect is also found true for unidirectional coupling when the central oscillator drives the identical outer oscillators. A consequence of this observation is that a mismatched central oscillator can drive many identical oscillators in a star-like configuration into enhanced synchrony. Further, in a ring of coupled oscillators, the enhancing is seen in oscillators in symmetric positions to a mismatched oscillator. The effect appears to be a general feature of nonlinear dynamical systems since a parameter regime of LS scenario [6] can always be found in chaotic systems. We are currently investigating details of the LS scenario in both limit cycle and chaotic regimes of coupled oscillators for further understanding of the enhancing effect.

This work is partially supported by the BRNS/DAE, India (#2009/34/26/BRNS). L. M. Pecora is supported by the Naval Research Laboratory, Washington.

___________________